\documentclass[12pt]{article}
\usepackage{graphicx}

\title{Confinement and $\alpha_s$  in a strong magnetic field}
\author{  Yu.A.Simonov, M.A.Trusov \\ Institute of Theoretical and Experimental
Physics\\ 117218, Moscow, B.Cheremushkinskaya 25, Russia}
\date{}

\newcommand{\be}{\begin{equation}}
\newcommand{\ee}{\end{equation}}

\def\la{\mathrel{\mathpalette\fun <}}

\def\fun#1#2{\lower3.6pt\vbox{\baselineskip0pt\lineskip.9pt
\ialign{$\mathsurround=0pt#1\hfil ##\hfil$\crcr#2\crcr\sim\crcr}}}

\newcommand{{\SD}}{\rm SD}

\newcommand{\vex}{\mbox{\boldmath${\rm x}$}}
\newcommand{\vey}{\mbox{\boldmath${\rm y}$}}
\newcommand{\ver}{\mbox{\boldmath${\rm r}$}}
\newcommand{\vesig}{\mbox{\boldmath${\rm \sigma}$}}

\newcommand{\veP}{\mbox{\boldmath${\rm P}$}}
\newcommand{\vep}{\mbox{\boldmath${\rm p}$}}
\newcommand{\veq}{\mbox{\boldmath${\rm q}$}}

\newcommand{\veL}{\mbox{\boldmath${\rm L}$}}

\newcommand{\veR}{\mbox{\boldmath${\rm R}$}}
\newcommand{\veA}{\mbox{\boldmath${\rm A}$}}

\newcommand{\ven}{\mbox{\boldmath${\rm n}$}}

\newcommand{\verho}{\mbox{\boldmath${\rm \rho}$}}

\newcommand{\veta}{\mbox{\boldmath${\rm \eta}$}}
\newcommand{\veB}{\mbox{\boldmath${\rm B}$}}

\newcommand{\vepi}{\mbox{\boldmath${\rm \pi}$}}

\newcommand{{\Mc}}{\mathcal{M}}

\newcommand{\lan}{\langle}
\newcommand{\ran}{\rangle}

\newcommand{\veBc}{\mbox{\boldmath${\rm \mathcal{B}}$}}
\newcommand{\veEc}{\mbox{\boldmath${\rm \mathcal{E}}$}}

\begin{document}

\maketitle
\begin{abstract}
Hadron decay widths are shown to increase in strong magnetic fields as $\Gamma (eB) \sim \frac{eB}{\kappa}
\Gamma(0)$. The same  mechanism is shown to be present  in the production of the sea quark pair inside the
confining string, which decreases the string tension with the growing  $eB$ parallel to the string . On the other
hand, the average energy  of the $q\bar q$ holes in the string world  sheet increases, when the direction of {\bf
B} is perpendicular to the sheet. These two effects stipulate the spectacular picture of the {\bf B} dependent
confinement and $\alpha_s$, discovered on the lattice.
\end{abstract}

\section{}
The QCD confinement (as well as perturbative  gluon exchange) was shown to be
created by the nonperturbative (np) color-electric field correlators
\cite{1,2,3} which are not  affected by magnetic field (m.f.) in the lowest
order in $\alpha_s$. However, in the next order in $\alpha_s$ (or in the
$1/N_c$ expansion) both confinement  and  gluon exchange (GE) interaction
contain quark loops, which interact with the m.f. and can influence the
resulting potentials.

For the GE part it was found in \cite{4}, that the  energy growth   of the
virtual  $q\bar q$ in m.f. prevents the original $Q\bar Q$ system from the
collapse, keeping the GE interaction $\lan V_{\rm GE} (\veq)\ran $ finite at
all $eB$.

An  interesting picture has emerged from the recent lattice studies in
\cite{5}, where it was shown, that confinement interaction $V_{\rm conf} (R)$
decreases for $\veB$ parallel to  $\veR$ and increases for the perpendicular
orientation, while $|V_{GE}|$ behaves in the opposite way. In the present paper
we suggest an explanation of the behavior $V_{conf} (R)$ and $\alpha_s$, and
simultaneously we point out the stimulating role of m.f. in the strong hadron
decay process.

The paper is organized as follows. In the next section we display  the path
integral Hamiltonian in m.f., the resulting wave functions, and some properties
of the spectrum  for the opposite charge  $q\bar q$ systems. In section 3 we
derive shortly the magnetic focusing effect  in  the creation of the $q\bar q$
pair. In section 4 we describe the appearance of sea quark holes in the
confining film  and the m.f. dependence of the  resulting effective string
tension. We also discuss the   dependence on the relative direction of  m.f.
and  make a comparison with lattice data.  In section 5  the $\alpha_s$
dependence on m.f. is derived and compared to lattice data.  In section 6
 we compare our results with the effective action expansion and lattice data
  on average field strength  squared in m.f. In section 7 a quantitative
  comparison of our results with the lattice data is presented.
   Section 8
   is devoted to the summary of results and possible developments of the
effects presented in the paper.

\section{Hamiltonian technic for hadrons in magnetic field}

In this section we exploit the path-integral Hamiltonian approach for the
$q\bar q$ systems (mesons) in m.f., which for the neutral case is embodied in
the Hamiltonian \cite{6,7,8}  \be H= \frac{\veP^2}{2(\omega_1+ \omega_2)} +
\frac{\vepi^2}{2\tilde \omega} + U (\veta) + \sum_{i=1,2} \frac{m_i^2+
\omega_i^2}{2\omega_i},\label{1}\ee where $\veP$ is the total momentum, $\vepi=
\frac{1}{i} \frac{\partial}{\partial\veta}$, and $\veta$ is the relative $q\bar
q$ distance, while $m_1 = m_2 \equiv   m$ is the current quark mass. Here
$\omega_i$ is the (virtual) energy of the quark $i$ which should be found from
the minimum of the total energy eigenvalue, $H\Psi_n = E_n \Psi_n$ and \be
\left.\frac{\partial E_n (\omega_1, \omega_2)}{\partial
\omega_i}\right|_{\omega_i = \omega_i^{(0)}}=0.\label{2}\ee The resulting
stationary value $E_n(\omega_1^{(0)}, \omega_2^{(0)})$ is the actual energy of
the $q\bar q$ system.

In the course of the decay the $q\bar q$ pair appears nearby the string connecting the original quarks $Q$ and
$\bar Q$, which are assumed to be heavy for simplicity.  We  shall show in this section,  that the magnetic
focusing effect \cite{9} is acting since both $q$ and $\bar q$ are charged. In this case (ignoring the c.m.
motion, $\veP\equiv 0)$ one should write the total Hamiltonian as \be H_{Q\bar Q}^{(q\bar q)} =
\frac{\vep^2_1+m^2_1}{2\omega_1} + \frac{\vep^2_2+m^2_2}{2\omega_2} +\frac{\omega_1+\omega_2}{2} + U(\ver_1 -
\veR_{\bar Q}) +U(\ver_2 - \veR_{ Q}) .\label{3}\ee The solution is readily obtains as a sum of two heavy-light
mesons, centered at $\veR_Q$ and $\veR_{\bar Q}$. However, one should impose the condition of the relative state
quantum number for $q\bar q$ which can be created by the  nonperturbative (n.p.) or perturbative mechanism,
yielding $J^{PC} = 0^{++} (^3P_0$ mechanism) or    $1^{--}$ ($^3S_1$ mechanism)  $q\bar q$ states respectively.

Now let us switch on the m.f. The Hamiltonian (\ref{3}) transforms as follows
\be H_{Q\bar Q}^{(q\bar q)}(B) = \sum_{i=q,\bar q} \frac{(\vep_i^\bot - e_i
\veA)^2 + \omega^2_i + m^2_i - e_i \vesig^{(i)} \veB +
(p_i^{(\|)})^2}{2\omega_i}+ U(\ver_1 - \veR_{ \bar Q} ) + U(\ver_2 - \veR_{ Q}
).\label{4}\ee Here $\veA (\ver) = \frac12 (\veB\times \ver)$, and it is
convenient to choose the origin $\ver=0$ just in the middle of the distance
$(\veR_Q- \veR_{\bar Q})$. Now one can separate out the center of mass motion
using the coordinates \be \verho =\frac{\omega_1 \ver_1 + \omega_2
\ver_2}{\omega_1 + \omega_2}, ~~ \veta = \ver_1 -\ver_2,~~ \vepi = \frac{1}{i}
\frac{\partial}{\partial \veta}, ~~\veP =\frac{1}{i} \frac{\partial}{\partial
\verho},\label{5}\ee and one has \be H_{QQ}^{(q\bar q)} (B) = H(\veP) + H_\pi +
U\label{6}\ee where $H(\veP)$ can be eliminated using the pseudomomentum
procedure as in \cite{8}, $U$ stands for  the last two terms  in (\ref{4}), and
\be H_\pi = \frac{\vepi^2}{2\tilde \omega} + \sum_{i=1,2}
\frac{m^2_i+\omega^2_i - e_i \vesig_i \veB}{2\omega_i} + \frac{\tilde \omega
\Omega^2_\eta \veta^2_\bot}{2} + X_\eta \veL_\eta \veB, \label{7}\ee where
\cite{10}, $ (e_1=e=-e_2)$ and subscripts $\bot$ and $\|$ refer to the
direction of $\veB$,  \be X_\eta = - \frac{e(\omega_2 -
\omega_1)}{2\omega_1\omega_2}, ~~ \Omega_\eta = \frac{eB}{2\tilde
\omega}.\label{8}\ee We take into account, that expanding $U$ in powers of
$\ver_i$ one has $U=\sigma R - \sigma\veta\ven +O(r^2_i)$ where $\veR =
\veR_{\bar Q} - \veR_Q, \ven=\frac{\veR}{R}$, and therefore  disregarding $U$
in the first approximation, one has a solution for $H_\pi (\omega_1 =\omega_2
=\omega) $\be M(\omega) = \frac{m^2+\omega^2}{\omega} + \frac{eB}{\omega}
(2n_\bot+1),\label{10}\ee yielding at the stationary point $\left.
\frac{\partial M}{\partial\omega}\right|_{\omega= \omega_0}=0$, \be M_0 \equiv
M(\omega_0) =2 \sqrt{m^2+e_qB}.\label{11}\ee Note, that in both cases $^3P_0
(S=1, L=1)$  and $^3S_1$ the spin  and orbital projections cancel. Hence the
$q\bar q$ pair acquires the effective mass (\ref{11}), which grows with $eB$,
when the  $q\bar q$ loop stays in the confining film,   when $\veB$ is
perpendicular to $\veR$.

 One can easily see in (\ref{7}), (\ref{8}) that the situation is different in
 the  case when $\veB$ is parallel to the  $q\bar q$ loop trajectory  since in
 this case $\Omega_\eta =0$ in (\ref{8}) and the  resulting $M_0=2m$ in
 (\ref{11}).

However for the transverse m.f.  the string  acquires  additional energy $M_0,
$ Eq. (\ref{11}), and   the total energy of the $Q\bar Q$ string with the
$q\bar q$ hole can be estimated as  $E(R) = V_{Q\bar Q} (R) + M_0 = \sigma R
+M_0$, and the resulting ratio of the  energy increase per one hole is \be
\frac{\Delta E(B)}{E(R)} \cong \frac{ 2\sqrt{ m^2_q + e_qB}}{\sigma R}
\label{11b}\ee

\section{Magnetic focusing  in  the $q\bar q$ pair creation}

Magnetic focusing was treated  in  \cite{9} in the case of two elementary
objects;  we now take the case of hadron constituents in (\ref{4}). Consider
the expansion of $U$ in the  powers at the ratios $\frac{r_i}{R} , ~~ i= 1,2$.
 Taking $\verho = \frac{\ver_1+\ver_2}{2}$, one has \be U=\sigma R -
\sigma \eta_\| + \frac{\sigma}{R} (2\rho^2_\bot +
\frac12\veta^2),\label{11a}\ee where the subscripts $(\|)$ and $(\bot)$ stand
for parallel and perpendicular with respect to $\veR$.
 Taking into account (\ref{7}), and solving $H\varphi =  (H_\pi +U) \varphi$,  one obtains
  the $B$-dependent
wave function  (for $n_\bot =0)$\be \varphi_0 (\eta_\bot) = \frac{\exp
(-\eta^2_\bot/2r^2_\bot)}{\sqrt{\pi} r_\bot} , ~~\frac{1}{  r^2_\bot} =
\sqrt{\frac{2\sigma \omega_0}{R} + (e_qB)^2} ,\label{13}\ee where $\omega_0$ is
to be  found from the stationary point of the  $\omega$-dependent energy, as in
(\ref{2}).
 For the lowest energy state one has from (\ref{13}) and (\ref{7}).

\be E(\omega) = \omega+\frac12 \sqrt{\frac{2\sigma}{\omega R}} +
\frac{1}{\omega} \sqrt{\frac{2\sigma \omega}{R} + (e_qB)^2}.\label{14a}\ee

Taking the minimum of (\ref{14a}), one finds $r^2_\bot$ and hence
$\varphi_0(0)$. Now the $q\bar q$ pair creation is described by the $q\bar q$
Green's function \cite{6} (in the background of the original $Q\bar Q$ Wilson
loop),$G_{q\bar q} (x,y) \sim \lan \vex,\vex|e^{-H(x_4-y_4)}|\vey\vey\ran \sim
|\varphi (0)|^2 e^{-ET}.$
 Hence the change in the wave function due to $B$ can be characterized by the
magnetic focusing factor \be \xi= \frac{\varphi^2_0 (\eta_\bot
=0;eB)}{\varphi^2_0 (\eta_\bot =0;0)} = \frac{\sqrt{\frac{2\sigma \omega_0
(eB)}{R} +(e_qB)^2}}{\sqrt{\frac{2\sigma \omega_0 (0)}{R}}} ,\label{14}\ee
 One can
see in (\ref{14}) two limiting cases \be a)~~ e_qB \ll \kappa^2, ~~ \xi\cong 1+
\frac{(e_qB)^2}{\kappa^4} ,~~ \kappa^2 = \left(
\frac{\sqrt{3}\sigma}{R}\right)^{2/3}\label{15}\ee

\be b)~~ e_qB \gg \kappa^2, ~~ \xi\approx \frac{e_qB}{\kappa^2}.\label{16}\ee

For $R\approx  1 $fm one has $\kappa^2=0.14$ GeV$^2$  ($\kappa^2 =0.22$ Gev$^2$ for $R=0.5$ fm), and one obtains
a strong amplifying factor for $\veB\|\veR$ and $eB\approx 1 $ GeV$^2$.

\section{Sea quark effects in the confinement regime}

It is clear, that m.f. acts on the fixed boundary Wilson loop $W_{Q\bar Q} (A)$ through the creation of sea quark
loops, which effectively create the holes in the film, covering the original Wilson  loop.

Following \cite{11, 12} one can write the partition function with the account
of sea quark loops as \be Z=\int DA \exp \mathcal{L}_A W_{Q\bar Q} (A) det (m_q
+ \hat D (A).\label{17}\ee where $det (m_q + \hat D(A)$ can be written in the
path integral form \be det (m_q + \hat D (A) ) = \exp \left[ tr  \left(
-\frac{1}{2} \int^\infty_0 \frac{ds}{s} \left( D^4z\right) e^{-K(s)}
W_{q\bar{q}}
 (A) \right) \right].\label{18}\ee
 Here $W_{q\bar q}$ is the closed loop of the sea quark and \be K(s) = \frac14
 \int^s_0 \left( \frac{dz_\mu (\tau)}{d\tau}\right)^2 d \tau + m^2_q
 s.\label{19}
 \ee
 Expanding (\ref{18}) in powers of $W_{q\bar q}$ and averaging  over DA, one
 obtains the effective one-loop partition function \cite{11, 12}
 \be Z_{1 loop} = - \frac12 \int^\infty_0 \frac{ds}{s} (D^4 z)_{xx} d^4x ~e^{-K(s)} \chi
 (W_{q\bar q}, W_{Q\bar Q} ), \label{20}\ee
 where $\chi$ is a  a connected average of the product of two loops
 \be \chi = \lan W_{q\bar q} (A) W_{Q\bar Q} (A) \ran - \lan W_{q\bar q} (A)
 \ran \lan W_{Q\bar Q}  (A) \ran. \label{21}\ee

 The properties of $\chi$ for different contour  orientations of  $C_{q\bar q}$ and $C_{Q\bar Q}$
 have been studied in \cite{13,14,15}, and in \cite{14} it was found , that for the
 simplest case of the flat overlapping contours of opposite orientation  one  can approximate $\chi$ as
 follows
 \be \chi\approx \frac{1}{N^2_c} \exp (- \sigma S_\Delta) = \frac{1}{N^2_c} \exp (- \sigma
 _{ren}S)\label{22}\ee
 where $S_\Delta$ is the   area  with subtracted area of loops  $q\bar q$,
    and $\sigma_{ren} = \left\lan \frac{S_\Delta}{S}\right\ran\sigma $ is  the
 string tension renormalized with account of sea quarks holes.

We define the density of the sea quark holes in the confining film in $W_{Q\bar
Q}(R,T)$, ~~ $\rho = \frac{\Delta S}{S}$, where $\Delta S$ is the area of the
holes, in the case of zero m.f., and follow the development of $\rho$ with the
magnetic field. It is clear, that the increasing energy of the holes yields the
increase of the effective string tension, which can be estimated  from
(\ref{11b}) as \be \frac{\Delta \sigma (eB)}{\sigma} = \rho \frac{\Delta
E(B)}{E(R)} = \frac{\Delta S}{S} \frac{2 \sqrt{e_qB}}{\sigma R}, \label{23}\ee
while the growth of $\rho$ due to magnetic focusing in the case of $B_\|$
should decrease effective string tension with $\Delta \sigma \sim \rho \xi
(eB)$.

As a result one can write,  taking into account, that \be \frac{\Delta\sigma
(eB=0)}{\sigma} = \frac{\Delta S}{S} = \rho, ~~ \frac{\overline{\Delta \sigma
(eB)}}{\sigma} = \frac{\Delta \sigma (eB) - \Delta \sigma
(eB=0)}{\sigma}\label{24}\ee

 \be\frac{ \overline{\Delta \sigma (eB) }}{\sigma} = \frac{\Delta S}{S} (f_\bot (eB) - f_\|(eB)),
  \label{25a}\ee where

\be f_\bot (eB) = \frac{2\sqrt{|e_q B|}}{\sigma R} , \label{26a}\ee and for $f_\| (eB)$ one has only the magnetic
focusing effect, \be f_\| (eB) = \xi (eB) {-1} .\label{27a}\ee

Note,  that the signs of both terms (\ref{26a}), (\ref{27a}) in  Eq.
(\ref{25a}), are opposite.

 One must have in mind, that  the  term $f_\| (eB)$ is present for the parallel
 direction of the m.f., $\veB= \veB\|$,  while the second term on the  r.h.s. in (\ref{25a})
 $f_\bot(eB)$  is active  when
 magnetic field is perpendicular  to the area.

\section{ Perturbative gluon exchange in magnetic field}

We now turn to the gluon exchange interaction $V_{OGE}$ in  magnetic field,
which was studied on the lattice in \cite{5,16,17} and analytically in
\cite{18}, and exploited in \cite{19} to predict the meson mass behavior in
m.f.

It was argued in \cite{18}, that m.f. creates a screening effect in $V_{OGE}$
due to the appearance of the quark loop contribution, which grows in m.f. in
the same way, as the quark pair energy (\ref{11}). This effect  was known  for
a long time  \cite{20} and  was exploited  in \cite{21} to predict the
saturating effect in QED.  Following this line  in the framework of QCD in
\cite{18} was obtained the one-loop $V_{OGE}$  with the dependence on m.f. in
the form

\be V_{OGE} (Q) = - \frac{16\pi}{Q^2} \frac{ \bar \alpha_s}{\left( 1+ \frac{
\bar \alpha_s n_f |e_q B|}{\pi Q^2}\exp \left( \frac{ - q^2_\bot}{2|e_q
B|}\right) T \left( \frac{q^2_\|}{4\sigma}\right)\right)}, \label{28a}\ee where
$T(z) \cong \frac{ 2z}{3+2z}$, $Q^2 = q^2_\bot + q^2_\|$, and \be \bar \alpha_s
= \frac{ \alpha_s^{(0)}}{1+ \frac{\alpha_s^{(0)}}{4\pi} \bar \beta \ln \frac{
Q^2 + M^2_B}{M^2_0}}= \frac{4\pi}{\bar{ \beta} ln
\frac{Q^2+M^2_B}{\Lambda^2_{QCD}}};
 ~~\bar \beta = \frac{11}{3}
N_c.\label{29a}\ee

When one is  measuring $V_{OGE} (R)$ on the lattice with $R\sim (0.5 \div 1)$ fm, one has $q^2_\|< \sigma$ and
small or vanishing $q_\bot$. Correspondingly one can expand $T(z)$ and  rewrite (\ref{28a}) as \be \lan V_{OGE}
(R) \ran_{eB} = \lan V_{OGE} (R) \ran_0 \frac{ \bar \alpha_s}{1+ C \bar \alpha_s}, ~~ C= \frac{ n_{f
}|e_qB|}{6\pi \sigma} \label{30a}\ee and \be \frac{ \Delta \alpha (eB)}{\bar \alpha_s } = -\bar \alpha_s + \frac{
\bar \alpha_s}{1+ C\bar \alpha_s} = -\frac{C\bar \alpha_s}{1+ C \bar \alpha_s}.\label{31a}\ee


Note the difference between the screening situation in QCD and QED. In QED
there is no string, and hence no string direction $\veR$, and the exchange and
the  $e^+e^-$, loops, transverse with respect to $\veB$, become heavy $(\sim
\sqrt{eB}$, Eq. (\ref{10})) and this effect screens  the Coulomb interaction in
the transverse direction.

In QCD the confining film (the string) defines the direction $\veR$, with the
sea quark loop lying inside the  film and hence one should have the screening
effect as in (\ref{31a}) for $\veR \bot\veB$ and no screening in the case
$\veR\|\veB$, when sea quarks move in  the loops along m.f.

In this case, however, the focusing effect, $\xi (eB) >1$, is acting,
increasing the sea quark loop density $\rho = \frac{\Delta S}{S}$ as
$\frac{\Delta S}{S} (\xi(eB)-1)$. This density is entering the general one-loop
expression (\ref{29a}) for $\alpha_s$, where $\bar \beta = \beta_0 =
\frac{11}{3} N_c - \frac23 n_f$ and the last two factors estimate the relative
density of gluon and quark loops respectively. In our case the increased
density of quark loops leads to the replacement in (\ref{29a})

\be \bar \beta \to \beta_0 + \Delta \beta_0 = \frac{11}{3} N_c - \frac23 n_f -
\frac29 n_f  (\xi (eB)-1),\label{33}\ee since only 1/3 the $n_f$ quark loops
lies in the parallel to $\veB$ position.

 Expanding in (\ref{29a}) in powers of
$(\xi-1)$, one obtains \be\frac{ \Delta \alpha_s}{\alpha_s} = \frac29
\frac{n_f}{\beta_0} (\xi-1).\label{34}\ee

One can see different signs of the m.f. action on $  \alpha_s$ in   (\ref{31a})
and (\ref{34}).
\section{Comparison to the effective action expansion}

We now turn to the general arguments, based on the expansion of the effective action $S_{eff}$, corresponding to
(\ref{18}), namely  we define as in \cite{17}, appendix D, \be \left\langle\det (m_q + \hat D(A))\right\rangle =
\exp (-S_{eff}).\label{6.1}\ee

The fourth order term in the expansion  of $S_{eff}$ in powers of constant
field  terms was obtained in \cite{22} and generalized to the case of the
superposition of magnetic field $B$ and colorelecric field $\veEc$ and
colormagnetic $\veBc$ in \cite{17}.

The $O(B^2)$ contribution has  the form (see Eq. (D.5) from \cite{17}) \be
S_{eff}^{(2,2)} =- \frac{V_4}{180 \pi^2} \frac{(eB)^2}{m^4_q} [ 3 tr \veBc^2_\|
+ tr \veBc^2_\bot + tr \veEc_\bot - \frac52 tr \veEc^2_\|].\label{6.2}\ee

Now taking into account, that the partition function $Z$ (\ref{18}) is
proportional to $\exp (-S_{eff})$, one can immediately see, that in the case
$\veEc\bot \veB$ (i.e. $\veB$ orthogonal to the Wilson loop surface, which was
denoted above in the paper as the case of $\veB_\bot$), one has \be \exp
(-S_{eff} ) = \exp (|{\rm const} | (eB)^2\veEc^2_\bot)>1,\label{6.3}\ee while
in the case of $\veB_\|$ one obtains \be \exp (-S_{eff} ) = \exp (-|{\rm const}
| (eB)^2\veEc^2_\|)<1.\label{6.4}\ee

The string tension is obtained from the correlator of the colorelectric fields
\cite{1} \be \sigma = \frac12 \int D(x) d^2 x, ~~ D(x,y) \sim \lan tr \veEc_i
(x) \phi  \veEc_i(y) \phi\ran .\label{6.5}\ee

Note, that $\exp (-S_{eff})$ enters as a factor in the field averaging denoted
by angular brackets (\ref{6.5}). Hence Eqs. (\ref{6.3}), (\ref{6.4}) tell us,
that

\be \sigma (\veEc_\bot, \veB ) > \sigma (0,0)\label{6.6}\ee

\be \sigma (\veEc_\|, \veB ) < \sigma (0,0)\label{6.7}\ee in agreement with
lattice measurements of \cite{5}.

On the lattice  the relevant behavior for $\lan \veEc^2_\|\ran$ and $\lan
\veEc^2_\bot\ran$ was found first in \cite{22} for the $SU(2)$ group and in
\cite{ 17} for the real QCD and is in  agreement with \cite{5} and our results
for $\Delta \sigma (B_\|)$ and $\Delta \sigma (B_\bot)$ respectively.

Note also, that in (\ref{6.3}), (\ref{6.4}) the coefficient of $\veEc^2_\|$ is
2.5 times bigger than that of $\veEc^2_\bot$, which is  qualitatively similar
to our relations of $\Delta \sigma (B_\|)$ and $\Delta \sigma (B_\bot)$,
following from $f_\|$ and $f_\bot$, Eqs. (\ref{27a}) and (\ref{26a}).

\section{Comparison  to the  lattice data  [5]}


To compare with  numerical data one should fix the parameters,  entering in our equations
(\ref{25a})-(\ref{27a}), (\ref{31a}), (\ref{34}).  Actually, the relative density of the $q\bar q$ holes in the
confinement area $\frac{\Delta S}{S}$ is the only free parameter of our approach and we choose it as
$\frac{\Delta S}{S}=0.15$, i.e. we suggest that the holes of sea quark loops occupy $\sim 15\%$ of the whole area
in absence of m.f.

 For  $f_\bot$ (\ref{26a}) and $\xi (eB)$, Eq. (\ref{15}) one should define the
 average value of $\lan \frac{1}{R}\ran$ in the  lattice measurements, and we
 take it $  \lan\frac{1}{R}\ran = \frac{1}{0.75 ~{\rm fm}} = 0.267$ GeV, since a
 large part of measurement was done in the interval $0.5$ fm $< R <$~1~fm.
 Correspondingly, $\kappa^2 =0.16$ GeV$^2$ in (\ref{15}), and  one obtains
 \be \frac{\Delta \sigma (eB_\bot)}{\sigma} = 0.31 \sqrt{ \frac{eB}{1~{\rm
 GeV}^2}}, \label{42n}\ee
 where we have taken into account, that $\bar e_q B \cong \frac{1}{2} eB$ for
 $n_f =2+1$. Now  we turn to the function $\xi(eB)$ in  (\ref{14}), which  can
 be approximated as $\xi (eB) \cong \sqrt{1+ \frac{(e_q B)^2}{\kappa^4}}$, and
 again with $\bar e_q B \cong \frac12 eB$, one has
 \be \frac{\Delta \sigma (eB_\|)}{\sigma} = - \frac{\Delta s}{s} \left( \sqrt{1+ \frac14 \left(
  \frac{eB}{\kappa^2}\right)^2}-1\right). \label{43n}\ee

\begin{figure}
\includegraphics[width=100mm]{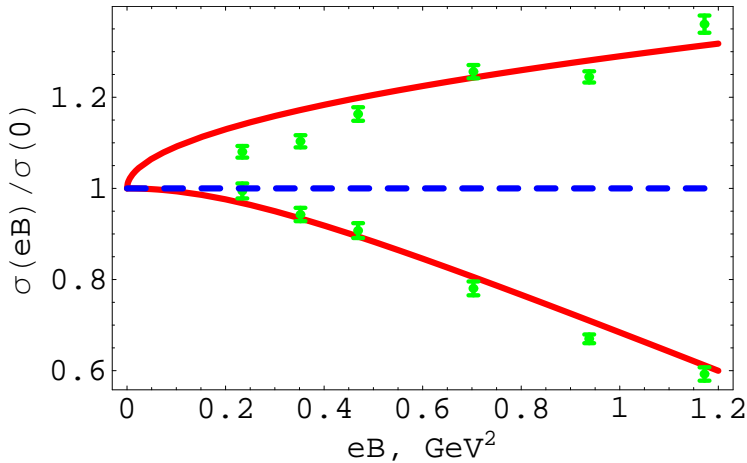}
\caption{$\sigma(eB)/\sigma(B=0)$ for $B_\bot$ (upper red curve) and $B_\|$ (lower red curve), in comparison with
lattice data from \cite{5} (green points)} \label{sigma_curve}
\end{figure}

The  resulting  curves of
 $\sigma (eB)$ for  $B_\bot$ and $B_\|$ are shown in Fig. \ref{sigma_curve} together  with the
lattice calculations of  \cite{5} (see Fig. 4 there at $L=40$). One can see a quantitative agreement in both
 cases with our estimate $\frac{\Delta S}{S} = 0.15$, and
 important agreement can be seen in the low $eB$ behavior, where (\ref{43n})
 yields quadratic growth $\delta \sigma (eB_\|) \sim - \frac{\Delta S}{S}
 \left(\frac{eB}{2\kappa^2}\right)^2$.

  We now turn to the  case of $\alpha_s (eB)$, Eqs. (\ref{30a})--(\ref{34}).
  In (\ref{30a}) one has for $n_f =3,$ $ \bar e_q \simeq e/2, $ $ C=0.44
  \frac{eB}{1~{\rm GeV}^2}$, and therefore with $\bar \alpha_s \cong 0.4$, one
  has

 \be \frac{\Delta \alpha_s(eB_\bot)}{\alpha_s} = -\frac{0.176(eB/{\rm GeV}^2)}{1+0.176(eB/{\rm GeV}^2)}
 , \label{44n}\ee
and  for $\alpha_s(eB_\|)$, writing (\ref{34}) to all orders of  $\xi^n$,

 \be \frac{\Delta \alpha_s(eB_\|)}{\alpha_s} \cong\frac{\sqrt{1+\left(\frac{eB}{2\kappa^2}\right)^2
 }-1}{14.5 -\sqrt{1+\left(\frac{eB}{2\kappa^2}\right)^2
 }} \label{45n}\ee

 It is clear that (\ref{45n}) is valid for $eB\la 1$ GeV$^2$.

%

\begin{figure}
\includegraphics[width=100mm]{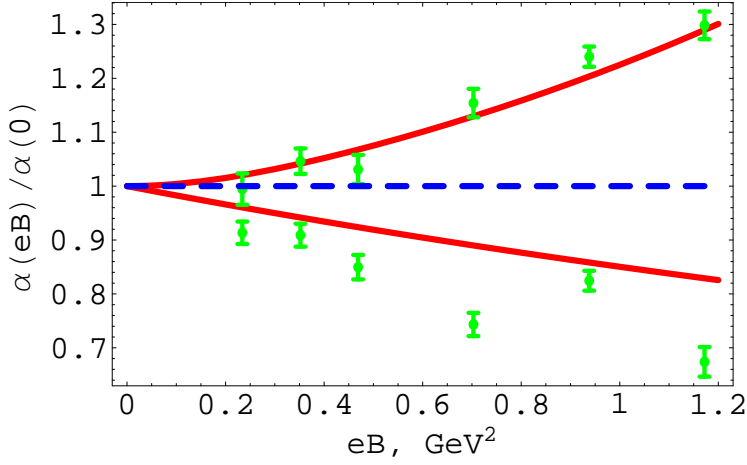}
\caption{$\alpha_s(eB)/\alpha_s(B=0)$ for $B_\bot$ (lower red curve) and $B_\|$ (upper red curve), in comparison
with lattice data from \cite{5} (green points)} \label{alpha_curve}
\end{figure}

The  resulting  curves of
 $\alpha (eB)$ for  $B_\bot$ and $B_\|$ are shown in Fig. \ref{alpha_curve} together  with the
lattice calculations of  \cite{5} (see Fig. 5 there at $L=40$). One can find the same type of behavior, and again
due to $\xi (eB)$, in our
 prediction for $\Delta \alpha_s (eB_\|)$ at small
 $\left(\frac{eB}{ \kappa^2}\right)$, which agrees well with the data of
 \cite{5}. Our values for $\Delta \alpha_s (eB_\bot)$ though are $\sim 40\%$ smaller in
 magnitude than the lattice data, but the general trend and sign are the same.
 Concluding, one can  notice a qualitative agreement and, for the case of $B_\|$,
 also a good quantitative agreement between our results and lattice
 measurements.

\section{Summary and conclusions}

Our discussion above is actually an attempt to qualitatively understand the dynamical mechanism beyond  the
$\sigma$ and $\alpha_s$ dependence on m.f. We have identified two possible effects in the action of external
magnetic field on confinement, which can act only through sea quarks loops. The first is the increasing
production of loops in m.f. -- the focusing effect. The second effect is  the energy increase due to loop
production, since they become effectively heavier in m.f., and this acts only when m.f. is perpendicular to the
area surface. As a result one obtains different signs of combining effects; as shown in Fig. \ref{sigma_curve}
and Fig. \ref{alpha_curve} and this corresponds to the lattice data \cite{5}.

 To make quantitative comparison with lattice data of \cite{5,16,17} the only
 fitting parameter is $\rho = \frac{\Delta S}{S}$ which was taken  as 0.15 . Note
 the difficulty in deriving it  from the general theory  \cite{1,2,3}, since the
 corresponding integrals are diverging and need regularization. The results for $\Delta\sigma$ are in a fine
 agreement with \cite{5}.

 We have calculated the screening of the $\alpha_s (eB)$ due to the quark pair
 creation in m.f., which occurs in $\veB_\bot$ and has the same physical
 mechanism as in the $q\bar q$ energy growth due to m.f., Eq. (\ref{11}).
  The stimulated creation of the $q\bar q$ pairs in the case of $B_\|$
 leads to the increase of $\alpha_s$.
  The resulting forms of $\Delta \alpha_s$ in Fig. \ref{alpha_curve} agree  with
 lattice data \cite{5},  and we have found the increase of $\alpha_s$ in $\veB_\|$
  due to the enhanced quark loop production, which is an antiscreening effect.

The authors are  grateful   to Massimo D'Elia for discussions, suggestions and numerical data. The authors are
grateful to M.A.Andreichikov, B.O.Kerbikov and members of the ITEP theory seminar  for useful discussions. The
financial support of the RFBR grant 1402-00395 is gratefully acknowledged.

%
%

\end{document}